\documentclass[aps,amsmath,amssymb,showpacs,showkeys]{revtex4}
\usepackage[dvips]{graphicx,color}
\usepackage{times}
\usepackage{multirow}
\usepackage{xcolor}
\usepackage[
  colorlinks=true,
  urlcolor=magenta,
  linkcolor=red,
  citecolor=blue]{hyperref}
\newcommand{\orcid}[1]{\href{https://orcid.org/#1}{\includegraphics[width=8pt]{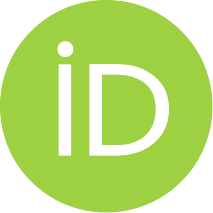}}}

\begin{document}
\title{Gravitational wave echoes from compact stars in $f(\mathcal{R},T)$ gravity}

\author{Jyatsnasree Bora \orcid{0000-0001-9751-5614}}
\email[Email: ]{jyatnasree.borah@gmail.com}

\affiliation{Department of Physics, Dibrugarh University,
Dibrugarh 786004, Assam, India}

\author{Umananda Dev Goswami \orcid{0000-0003-0012-7549}}
\email[Email: ]{umananda2@gmail.com}

\affiliation{Department of Physics, Dibrugarh University,
Dibrugarh 786004, Assam, India}

\begin{abstract}
We study the gravitational wave echoes from the static and spherically 
symmetric compact stars in $f(\mathcal{R},T)$ gravity metric formalism. In this
study, to describe the matter of compact stars we use the MIT Bag model and 
the color-flavor-locked (CFL) equations of state (EoSs). Solving the 
hydrostatic equilibrium equations i.e., the modified TOV equations in 
this gravity, we obtain different stellar models. The mass-radius profiles 
for such stellar configurations are eventually discussed. The stability 
of these configurations are then analysed using different model parameters. 
From the solutions of TOV equations, we check the compactness of such objects. 
It is found that similar to the unrealistic EoS, i.e.~the stiffer form of the 
MIT Bag model, under some considerations the realistic interacting quark 
matter CFL EoS can give stellar structures which are compact enough to possess 
a photon sphere outside the stellar boundary and hence can echo GWs. The 
obtained echo frequencies are found to lie in the range of 39-55 kHz. 
Moreover, we show that for different parametrizations of the gravity theory, 
the structure of stars and also the echo frequencies differ significantly. 
Further, we constrain the pairing constant value $\beta$ from the perspective 
of emission of echo frequencies. It shows that for the stiffer MIT Bag model 
$\beta\geq-2.474$ and for the CFL phase with massless quark condition 
$\beta\geq-0.873$, whereas for the massive case $\beta\geq-0.813$. 
\end{abstract}


\maketitle


\section{Introduction}\label{intro}
The theory of general relativity (GR) is the most alluring theory of gravity 
that could ever be formulated by the human mind. This theory works excellently 
in predicting various phenomena that Newtonian gravity fails to explain. Since 
the last 100 years it has been tested very successfully in the weak field limits. 
The decisive and most enthralling support to GR in the very strong field regime 
has come when the gravitational waves (GWs) have been detected by the LIGO 
detectors first in 2015 from the merging event of two black holes \cite{bpabott} and 
subsequent detections of such waves by these detectors \cite{currentligo} 
after the first detection. However, in spite these unparallel successes, 
GR has seen its unexpected limitations in the light of cosmological and 
astrophysical observations, such as the observed accelerated 
expansion of the universe \cite{Riess_1998, perl}, dark matter \cite{bertone} 
etc. and also it suffers from the theoretical front, such as it is not a 
renormalizable theory \cite{stelle}. These limitations suggest that GR is not an 
ultimate theory of gravity and hence it needs some modifications. Especially, in order 
to explain the current accelerated expansion of the universe and other observational 
issues \cite{grigorian,li} the modifications of GR were quite urgent. In such 
modifications the description of GR can be retained by applying necessary 
modifications in the Einstein-Hilbert (EH) action, which will result in the 
different alternative theories of gravity (ATGs). The most elementary and 
simplest way to extend GR is the $f(\mathcal{R})$ gravity theories 
\cite{sotiriou, felice} where, in the EH action the Ricci scalar $\mathcal{R}$ 
is replaced by the generic function of Ricci scalar i.e., $f(\mathcal{R})$. 
Another important and accepted modification to GR is the $f(\mathcal{R},T)$ 
gravity \cite{harko} in which the gravitational Lagrangian is represented as 
a function of the Ricci scalar $\mathcal{R}$ and of the trace of 
energy-momentum tensor $T$. Beside $f(\mathcal{R})$ and $f(\mathcal{R},T)$ gravity 
theories other such ATGs are $f(\mathcal{R},L_m)$ \cite{frlm}, 
$f(\mathcal{R},\square\mathcal{R})$ \cite{carloni}, $f(\mathcal{T})$ gravity \cite{ft} 
where $\mathcal{T}$ is the torsion scalar, Rastall gravity \cite{rastall}, etc. There 
are numbers of cosmological models proposed based on these ATGs and from such models 
an excellent agreement between theory and experiment can be obtained 
\cite{bahcall, hwang, demi, singh}.

As most of these ATGs show impressive agreement with the observational data 
in the low field regime (cosmological scale), it is necessary to test them in
the high gravity limits. This can be done by studying the massive compact
stellar objects, such as black holes, strange stars, neutron stars etc.~in ATGs
and compare the obtained properties of such objects with the experimentally 
observed ones. These sorts of studies have been carried out for the last many
years using different ATGs \cite{14stay,15oli,21pano,21geo,21nash,22pretel,22sil}. 
Specifically, the literature survey shows that over the last few years a plethora of 
articles are focusing on compact stars in the context of $f(\mathcal{R},T)$ gravity 
from various astrophysical points of view. Among many other references some important 
references in this direction are 
\cite{zubair,das,moraes,deb,yousaf,lobato,mustafa,pretel,pretel22}. D.~Deb et al.~studied a 
model of strange stars under the framework of $f(\mathcal{R},T)$ gravity and they 
examined the stability as well as the physical properties of the compact stellar 
system \cite{deb}. Neutron stars in $f(\mathcal{R},T)$ gravity are studied with the 
model, $\mathcal{R}+2\beta T$ by using some realistic hadronic equations of state 
(EoSs) in Ref.~\cite{lobato}. A constraint on the pairing constant $\beta$ is also 
reported therein. Considering the anisotropic fluid without electric charge 
for the stellar objects in $f(\mathcal{R},T)$ gravity, the stability 
conditions are reported in the Ref.\ \cite{mustafa}. Recently, the stability of 
compact stars in the $f(\mathcal{R},T)$ theory against radial perturbations 
are reported in \cite{pretel}. Using the standard MIT Bag model EoS, the 
radial perturbation equations are derived therein. This study demands that the 
stellar structures in $f(\mathcal{R},T)=\mathcal{R}+2\beta T$ theory are 
stable against radial perturbations, as for such theories lowest normal modes 
are found to be real \cite{pretel}. Again for the charged and 
anisotropic compact stars in the $f(\mathcal{R},T)$ theory, the stellar 
structures are analysed in the Ref.~\cite{rej}. 

The study of dense matter turns into an interesting and curious area of 
research after Witten's strange matter hypothesis \cite{witten}. It has 
paved a new way to think about compact dense matter. According to this 
hypothesis, strange matter is the true ground state of hadron \cite{witten}. 
Such matters are assumed to be composed of roughly equal numbers of $u$, $d$ 
and $s$ quarks and a negligible amount of electrons to maintain charge 
neutrality throughout its structure. To describe such matter, MIT Bag model is 
the simplest and most widely used EoS, where the quarks are in the deconfined 
or unpaired state. This EoS assumes a gas of free relativistic quarks and 
confinement is achieved through the Bag constant $B$. However at high density 
the attractive force among quarks that are antisymmetric in color tends to 
pair quarks that are near the Fermi surface. This pairing strength is 
controlled by the pairing gap parameter $\Delta$ and such pairings are the 
building blocks of color-flavor-locked (CFL) quark phase \cite{alford}. 
Recently, it has been shown that CFL state is more energetically favourable 
and it widens the stability window \cite{lugones}. Such a state can be 
potentially found in the dense core of neutron stars and thus forming hybrid 
stars \cite{alford1}. The study of such ultra-compact objects has an 
interesting observational prospect from the point of view of GWs in the sense 
that it will lead to validate different models of compact dense matter in 
future and hence may lead to ultimately understanding the compact stellar 
structures. This prospect is based on the fact that due to their high 
compactness as compared to other compact objects, such stars can echo the GWs 
falling on their surfaces \cite{2016_Cardoso,2016_Cardoso_prd,pani,2019_Cardoso,2021_Vlachos,2022_Chatzifotis}.
To acquire this interesting feature, compact stars 
should possess a photon sphere outside their stellar structures 
\cite{manarelli}. Such possibility of gravitational 
wave echoes (GWEs) from the ultra-compact stars were first proposed by P.\ Pani
and V.\ Ferrari \cite{pani}. For strange stars using the MIT Bag model EoS in 
the GR realm, GWEs are reported in the range of kHz in Ref.~\cite{manarelli}. 
The studies on the GWEs from ultra-compact stars in the context of GR with 
different EoS are reported in  Refs.~\cite{urbano, jb1, Zhang, jb2}. Recently 
in the realm of ATGs possibilities of echoes are reported in the 
Ref.~\cite{jb3} using the Palatini approach of $f(\mathcal{R})$ gravity 
considering three different models in the theory. Thus the GWEs can be the 
testing tool for the different models of ATGs also.  

Motivated from the above studies in this work we are interested to check 
echoes of GWs from the surface of ultra-compact stars, like strange stars in 
another ATG. So our present study involves compact stellar model in 
$f(\mathcal{R},T)$ gravity theory with one of its most promising model 
proposed by T.~Harko et al.~\cite{harko} as $f(\mathcal{R},T)=\mathcal{R}+2\beta T$, 
$\beta$ being a constant. In this work we have adopted the metric 
formalism of $f(\mathcal{R},T)$ gravity. Using this formalism compact stars 
are studied earlier in the $f(\mathcal{R},T)$ gravity in 
Refs.~\cite{lobato, pretel, rej}. Moreover the Palatini approach to this study in this 
theory can be found in Ref.~\cite{wu, barrientos, bhatti}. To the best of our 
knowledge the possibilities of GWEs in this theory have not yet been calculated. 
Also in this present study we have applied the realistic interacting quark 
matter EoS i.e., the CFL phase to construct strange star models. The stability 
of these constructed strange star models are also discussed briefly.

This present paper is organised as follows. The theory of $f(\mathcal{R},T)$ 
gravity is discussed shortly in the section \ref{frt}. A brief discussion on 
the modified TOV equations is also added to this section. In section \ref{eos}, 
the EoSs to describe compact matter are presented. The basics of GWEs and calculated 
results of these echoes for the model are discussed in the section \ref{gwe}. After 
this section the physical properties and related stability of the stellar 
configurations for some physical parameters of the model are briefly discussed in the 
section \ref{numerical}. Finally, we conclude our article in the section 
\ref{conclusion}. In this article, we adopt the unit of $c=G=1$, where $c$ and $G$ 
denote the speed of light and the gravitational constant respectively, and also we 
used the metric signature $(-,+,+,+)$.
 

\section{$f(\mathcal{R},T)$ gravity}\label{frt}
In the $f(\mathcal{R},T)$ theory of gravity proposed by T.~Harko et 
al.~\cite{harko} the modified EH action is an arbitrary function of 
$\mathcal{R}$ and $T$, the Ricci scalar and the trace of the energy-momentum 
tensor $T_{\mu\nu}$ respectively. Thus the action this gravity can be written 
as
\begin{equation}
\label{a}
S=\dfrac{1}{2\kappa}\int d^{4}x\sqrt{-g}f(\mathcal{R},T)+S_{m},
\end{equation}
where $\kappa\equiv 8\pi$ and $S_{m}$ is the action for the matter which 
depends on the metric $g_{\mu\nu}$ and the matter field $\Phi_{m}$. Varying 
this action \eqref{a} with respect to $g_{\mu\nu}$ will result in the field 
equation in the metric formalism as given by
\begin{equation}
\label{b}
f_{\mathcal{R}}(\mathcal{R},T)\mathcal{R}_{\mu\nu}-\dfrac{1}{2}g_{\mu\nu}f(\mathcal{R},T)+\big[g_{\mu\nu}\square-\nabla_{\mu}\nabla_{\nu}\big]f_{\mathcal{R}}(\mathcal{R},T)= \kappa T_{\mu\nu}-f_{T}(\mathcal{R},T)T_{\mu\nu}-f_{T}(\mathcal{R},T)\Theta_{\mu\nu}.
\end{equation}
In this equation $$f_{\mathcal{R}}(\mathcal{R},T)\equiv\dfrac{\partial
f(\mathcal{R},T)}{\partial \mathcal{R}},\;\; f_{T}(\mathcal{R},T)\equiv
\dfrac{\partial f(\mathcal{R},T)}{\partial T},\;\; \square\equiv\partial_{\mu}(\sqrt{-g}g^{\mu\nu}\partial_{\nu})/\sqrt{-g}$$ is the D'Alambert operator, 
$R_{\mu\nu }$ is the Ricci tensor, $\nabla_{\mu}$ represents the covariant 
derivative. The tensor $\Theta_{\mu\nu}$ is defined as
\begin{equation}\label{c}
\Theta_{\mu\nu}\equiv \dfrac{g^{\alpha\beta}\delta T_{\alpha\beta}}{\delta g^{\mu\nu}},
\end{equation} 
and the energy-momentum tensor $T_{\mu\nu}$ is defined as
\begin{equation}\label{d}
T_{\mu\nu}=g_{\mu\nu}\mathcal{L}_{m}-\dfrac{2\partial\mathcal{L}_{m}}{\partial g^{\mu\nu}},
\end{equation}
where $\mathcal{L}_{m}$ is the Lagrangian density for matter. 

Unlike the situation for GR where the energy-momentum tensor is a conserved 
quantity, in the $f(\mathcal{R},T)$ gravity the conservation condition for 
energy-momentum tensor is violated. This can be seen by taking the covariant 
derivative of equation \eqref{b}, which leads to the equation that violets 
conservation condition for $T_{\mu\nu}$ as
\begin{equation}
\label{e}
\nabla^{\mu}T_{\mu\nu}=\dfrac{f_{T}(\mathcal{R},T)}{\kappa-f_T(\mathcal{R},T)}\left[(T_{\mu\nu}+\Theta_{\mu\nu})\nabla^{\mu}\,\ln f_T(\mathcal{R},T)+\nabla^{\mu}\theta_{\mu\nu}-\dfrac{1}{2}g_{\mu\nu}\nabla^{\mu}T\right].
\end{equation} 
As mentioned earlier, in this work we have considered one of the most widely 
used $f(\mathcal{R},T)$ gravity model of the form $f(\mathcal{R},T)=
\mathcal{R}+2\beta T$. For this model, $f_{T}(\mathcal{R},T)=2\beta\neq 0$ 
and hence $\nabla^{\mu}T_{\mu\nu}\neq 0$. Thus this model is implying an energy 
non-conserving system. A detailed analysis in this regard can be found in 
Ref.~\cite{harko, deb}. From this model one can easily retrieve the 
predictions of GR just by substituting $\beta=0$. Again, in this particular 
model $\beta$ is the only free parameter and is known as the coupling 
constant. This parameter arises due to the coupling between matter and 
geometry in the modified gravity. However, it is still lacking proper 
constrained values for different astrophysical scenarios. Recently several 
literature demands some restrictions on the value of $\beta$ from some 
observations of astrophysical to cosmological scales. In Ref.~\cite{carvalho}, 
for massive white dwarfs using the observational data a lower bound on the 
model parameter $\beta$ was reported to be $\beta>-3\times 10^{-4}$. Another 
lower limit for $\beta$ was reported as $\beta\gtrsim -1.9\times 10^{-8}$ from 
dark energy density parameter in \cite{sahoo}. For the neutron star with crust,
R.~Lobato et al.~showed that $|\beta|\lesssim 0.02$ \cite{lobato}. From the 
cosmological context a constraint range of $\beta$ was reported as 
$-0.1<\beta<1.5$ \cite{velten}. Using the realistic EoSs for the compact 
stars, recently it is reported that $\beta<0$ \cite{lobato, pretel}. 

Now, in order to describe compact stars composed of the adiabatic and 
isotropic fluids we consider the energy-momentum tensor of perfect fluid as
\begin{equation}
\label{f}
T_{\mu\nu}=(p+\rho)u_{\mu}u_{\nu}+p\,g_{\mu \nu},
\end{equation}
where as usual $\rho$ represents the density, $p$ is the pressure of the 
isotropic fluid and $U_{\mu}$ are its four-velocities. The spherically 
symmetric and static metric for compact star can be written as
\begin{equation}
\label{g}
ds^{2} = -\,e^{\chi(r)}dt^{2}+e^{\lambda(r)}dr^{2}+r^{2}\,d\theta^{2}+r^{2}
	\sin^{2}\theta\,d\phi^{2}.
\end{equation}
The metric function $\chi$ and $\lambda$ are functions of radial coordinate 
$r$ only with the solutions,
\begin{equation}
\label{h}
e^{\chi(r)}=e^{-\lambda(r)}=1-\dfrac{2m(r)}{r}.
\end{equation}
Again, for the $f(\mathcal{R},T)=\mathcal{R}+2\beta T$ model of our interest 
the hydrostatic equilibrium equations or the Tolman-Oppenheimer-Volkoff (TOV) 
equations of isotropic fluid sphere can be found as \cite{moraes,pretel}
\begin{equation}
\label{i}
\dfrac{dm}{dr}=4\pi r^2 \rho+\dfrac{\beta}{2}r^2 (3\rho-p),
\end{equation}
\begin{equation}
\label{j}
\dfrac{dp}{dr}=-\,\dfrac{(\rho+p)}{1+a}\left[\dfrac{m}{r^2}+4\pi r p-\dfrac{\beta}{2}r(\rho-3p)\right]\left(1-\dfrac{2m}{r}\right)^{-1}+\dfrac{a}{1+a}\dfrac{d\rho}{dr},
\end{equation}
\begin{equation}
\label{k}
\dfrac{d\chi}{dr}=-\,\dfrac{2(1+a)}{\rho+p}\dfrac{dp}{dr}+\dfrac{2a}{\rho+p}\dfrac{d\rho}{dr}.
\end{equation} 
Here, the term $a$ is defined as 
\begin{equation}
\label{l}
a= \dfrac{\beta}{\kappa+2\beta}.
\end{equation}
It is to be noted that these modified TOV equations \eqref{i}--\eqref{k} 
become the usual TOV equations in GR when $\beta = 0.$ Moreover, in the case 
of the barotropic EoS, where $p = p(\rho)$, we can rewrite equation \eqref{j} 
as \cite{carvalho,pretel}
\begin{equation}
\label{addj}
\dfrac{dp}{dr}=-\,(\rho+p)\left[\dfrac{m}{r^2}+4\pi r p-\dfrac{\beta}{2}r(\rho-3p)\right]\left[1+a\left(1-\frac{d\rho}{dp}\right)\right]^{-1}\left(1-\dfrac{2m}{r}\right)^{-1}.
\end{equation}
This equation implies that the hydrostatic equilibrium configurations of the 
stellar structure will be possible only when
\begin{equation}
\label{m}
1+a\left(1-\frac{d\rho}{dp}\right)>0.
\end{equation}
This condition can be rearranged to a convenient form as
\begin{equation}
\label{addm}
a\left(1-\frac{dp}{d\rho}\right)<\frac{dp}{d\rho}.
\end{equation}
From this condition it is clear that since at the surface of the star 
$dp/d\rho\rightarrow 0$, we should have $a<0$. Eventually this leads to the
fact that the model parameter $\beta$ should be such that $\beta<0$. Thus in 
this work we will only consider the negative values of $\beta$. For the 
case of white dwarfs such a condition was earlier reported in the 
Ref.~\cite{carvalho} and for the strange stars the same condition was used in 
the Ref.~\cite{pretel}. 

The boundary conditions needed to solve the system of TOV equations is 
same as that of the GR case due to the linearity of $f(\mathcal{R},T)$ in 
$\mathcal{R}$. This will finally lead to the exterior solution as the 
Schwarzschild vacuum exterior solution. Therefore we have boundary 
conditions, at the centre of the star, $m=0$, $p=p_c$ and $\rho=\rho_c$ i.e., 
pressure and density take their respective central values. Whereas at the 
stellar surface i.e., at $r=R$, the surface pressure vanishes, $p=0$. At the 
surface of the star the interior solution connects with the Schwarzschild 
vacuum exterior solution. Thus the metric solutions at the surface are given 
by $e^{\chi(R)}=e^{-\lambda(R)}=1-2M/R$.

\section{Equations of state}\label{eos}
The solution of TOV equations can lead one to know some physical 
characteristic of compact stars such as the radius, mass etc.~only when these 
are supplemented with some relations between energy density and fluid pressure 
known as EoSs. In this study we shall use two EoSs to describe the dense 
matter. The first one is the simplest and most usually employed EoS namely the 
MIT Bag model EoS of the form \cite{witten},
\begin{equation}
\label{n}
p=\dfrac{1}{3}(\rho-4\,B). 
\end{equation}
This EoS describes deconfined quark matter composed of $u$, $d$ and $s$ quarks 
and the confinement pressure is obtained by the Bag constant $B$. One may note 
that solving the TOV equations together with the MIT Bag model will result in 
stellar configurations which are not compact enough to feature a photon sphere 
around their surface, irrespective of different $B$ values within its allowed 
range \cite{jb1}. However it is reported earlier that this EoS can be made a
stiffer one if we modify it as \cite{manarelli, jb1, jb2, jb3}
\begin{equation}
\label{o}
p=\rho-4\,B .
\end{equation}
In the present work we shall use this stiffer form of the MIT Bag model EoS. 
Also this EoS is good enough to have the desired range of compactness.

The other strange quark matter EoS we are interested in is the CFL phase EoS 
\cite{alford} as mentioned earlier. This CFL phase involves the formation of 
$ud$, $us$ and $ds$ Cooper pairs. The corresponding thermodynamic potential 
$\Omega_{CFL}$ of order $\Delta^2$ for this phase can be 
obtained as \cite{alford,lugones}
\begin{equation}
\label{p}
\Omega_{CFL}=\Omega_{free}-\dfrac{3 \Delta^2 \mu^2}{\pi^2} +B.
\end{equation}
The thermodynamic potential for the free quarks without pairing interaction 
is given by
\begin{eqnarray}
\label{q}
\Omega_{free}=\dfrac{6}{\pi^2}\int_{0}^{\nu}\big[p-\mu\big]p^2dp+\dfrac{3}{\pi^2}\int_{0}^{\nu}\left[(p^2+m_s^2)^{1/2}-\mu\right]p^2dp\\
=\sum_{i\,=\,u,d,s}\dfrac{1}{4\pi^2}\left[\mu_i \nu\left(\mu_i^2-\dfrac{5}{2}m_i^2\right)+\dfrac{3}{2}m_i^4 \log\left(\dfrac{\mu_i+\nu}{m_i}\right)\right],
\end{eqnarray}
where $\mu$ is the baryon chemical potential, $3\mu=\mu_u+\mu_d+\mu_s$ and 
$m_s$ is the mass of strange quark. As defined earlier $B$ is the Bag 
constant. Due to the pairing interaction forces the flavours have the same 
baryon number density $n_B$ and particle number densities,
\begin{equation}
\label{r}
n_B=n_u=n_d=n_s=\dfrac{(\nu^3+2\Delta^2\mu)}{\pi^2}
\end{equation}
and the common Fermi momentum is given by
\begin{eqnarray}
\nu=(\mu_i^2-m_i^2)^{1/2} = 2\mu-\left(\mu^2+\dfrac{m_s^2}{3}\right)^{1/2}\!\!\!.
\end{eqnarray}
In equation \eqref{p}, $\Delta$ is the pairing gap (the gap of the QCD Cooper 
pairs) which can be considered as a free parameter \cite{lugones} and the term 
$3 \Delta^2 \mu^2/\pi^2$ is the condensate term. The pressure and the 
energy density of the strange quark matter (SQM) can be obtained as follows:
\begin{equation}
\label{s}
p=-\,\Omega_{CFL},
\end{equation}
\begin{equation}
\label{t}
\rho=\sum_i \mu_i n_i+\Omega_{CFL}=3\,\mu\, n_B-p.
\end{equation}
This pressure-density relation for the SQM based on the CFL state can now be 
written as
\begin{equation}
\label{u}
\rho=3\,p+4B-\dfrac{9\,\alpha\,\mu^2}{\pi^2},
\end{equation}
which implies that
\begin{equation}
\label{v}
p=\dfrac{\rho}{3}-\dfrac{4B}{3}+\dfrac{3\,\alpha\,\mu^2}{\pi^2},
\end{equation}
where $\mu^2$ and $\alpha$ are given by
\begin{equation}
\label{w}
\mu^2=-\,3\,\alpha+\left[9\,\alpha^2+\dfrac{4}{3}\pi^2(p+B)\right]^{1/2}
=-\,\alpha+\left[\alpha^2+\dfrac{4}{9}\pi^2(\rho-B)\right]^{1/2}\!\!\!,
\end{equation}
and \begin{equation}
\label{y}
\alpha=-\dfrac{m_s^2}{6}+\dfrac{2\Delta^2}{3}.
\end{equation}
Lack of the accurate values of the parameters $\Delta$ and $m_s$ will allow 
us to consider these terms as free parameters and can be constrained using 
stability conditions \cite{maulana,flores,farhi}. In this study we have 
considered the massless quark case i.e., $m_s=0$ and a finite mass case 
$m_s\neq0$ considering $m_s=100$ MeV \cite{beringer}.


\section{Gravitational wave echoes}\label{gwe}
The understanding of black holes and compact stars has gained a new 
direction after the observations of GWs. The detection of binary black holes 
merging events inspired many researchers to look into the other exotic compact 
objects, which can act like black hole mimickers. Such exotic compact objects 
can maintain their existence with a high compactness without featuring an event 
horizon in contrast to the case of black holes. Due to their high compactness 
it was proposed earlier that such compact objects can generate echoes of the 
GWs falling on their gravitational potential barrier \cite{pani}. However, 
even though such compact objects do exist, they may develop some 
instabilities when rotating. Such instabilities of ultracompact stars are the 
ergoregion instability and non-linear instabilities as reported in Ref.~s 
\cite{1978_Friedman,1978_Comins,2008_Cardoso,2008_Chirenti,2014_Keir,2014_Cardoso,
2016_Moschidis,2022_Cunha}. Also, in such cases, the linear and non-linear stability is 
disturbed by the existence of very long-lived modes found in \cite{2014_Cardoso}. 
These real modes can trigger nonlinearities that cannot be probed with perturbative techniques. Such nonlinear instability of ultracompact spherically symmetric exotic 
objects is recently reported in the Ref.~\cite{2022_Boyanov}.
When GWs from a distant merging event falls on their surface it gets reflected at the 
photon sphere and after some time delay multiple reflection and refraction 
occurs. In order to generate echoes of GWs it is required that such objects 
should feature a photon sphere at $R_p=3M$, $M$ being the total mass of the 
star. Again, since one of the distinctive features of compact stars from black 
hole is the absence of an event horizon, so the minimum radius of such a star 
should be greater than the Buchdahl's radius $R_b=9/4 M$. It is to be pointed 
out that this Buchdahl's limit is applicable for the stars in the GR 
considerations only \cite{buchdahl} and for ATGs this value is modified as 
$R_b=9/(4-3c/2)M$ \cite{burikham}. Here $c=4\pi p_{eff}(R)R^2$, $p_{eff}(R)$ 
being the effective pressure at the stellar boundary. It is reported that the 
value of $c$ is negative and hence even for a small value of $c$ the 
Buchdahl's radius will decrease in the case of ATGs \cite{burikham}. 
Thus those compact stars whose radius lies in the limit $R_b \leq R \leq R_p$ 
are the promising candidates to echo the GWs falling on their stellar 
surfaces.

In order to calculate the echo frequencies, first the characteristic echo 
times are calculated by using the relation, 
\begin{equation}
	\label{z}
	\tau_{echo}\equiv\int_0^{3M}\!\!\!\! e^{\,(\lambda(r)-\chi(r))/2}\;dr.
	\end{equation} 
The metric functions $\lambda(r)$ and $\chi(r)$ are obtained from equation 
\eqref{h}. After calculating the echo time the echo frequencies can be 
calculated by using the relation $\omega_{echo} \approx \pi/\tau_{echo}$.

\begin{figure*}[!h]
        \centerline{
        \includegraphics[scale = 0.32]{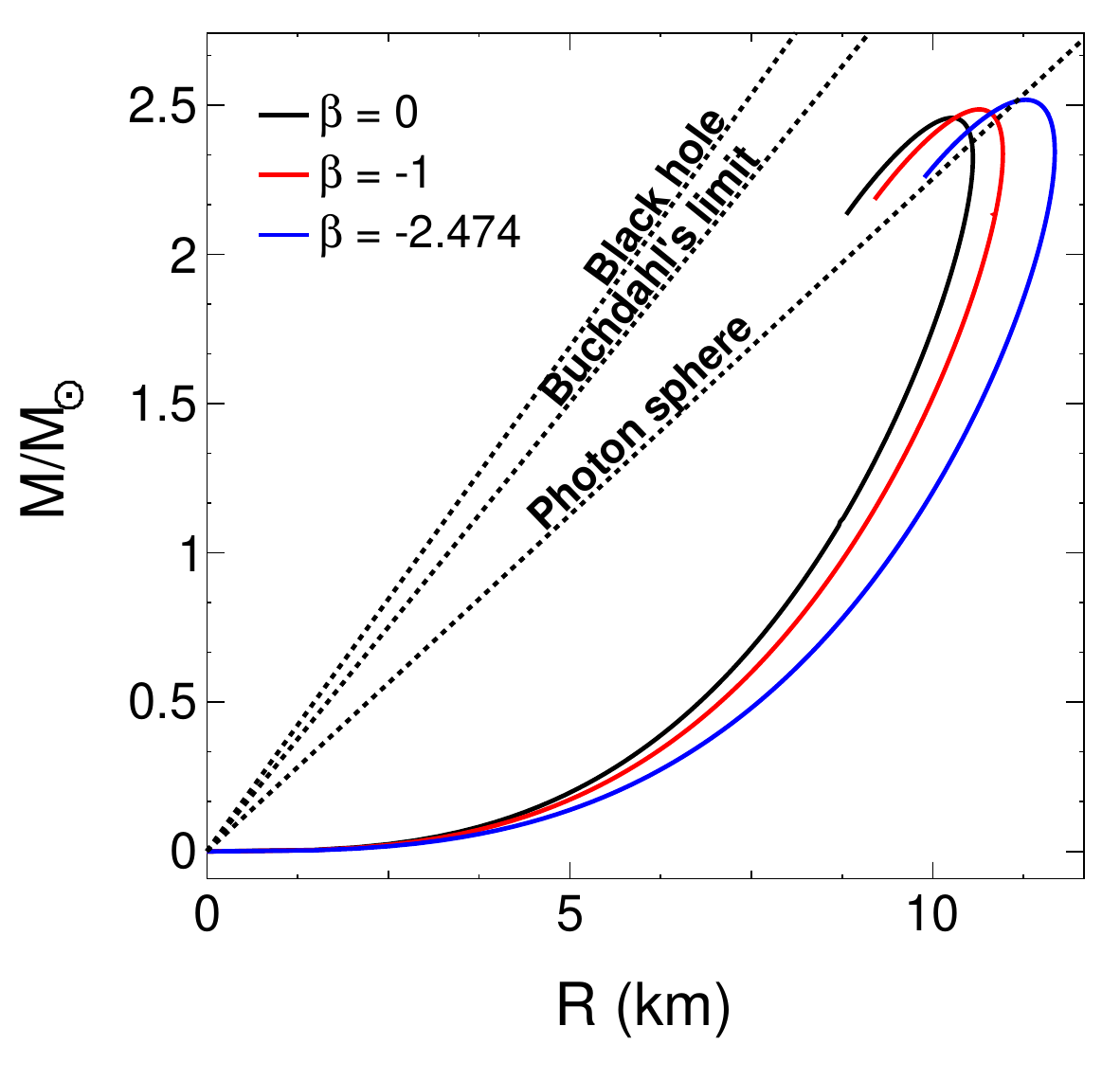}\hspace{1cm}
        \includegraphics[scale = 0.32]{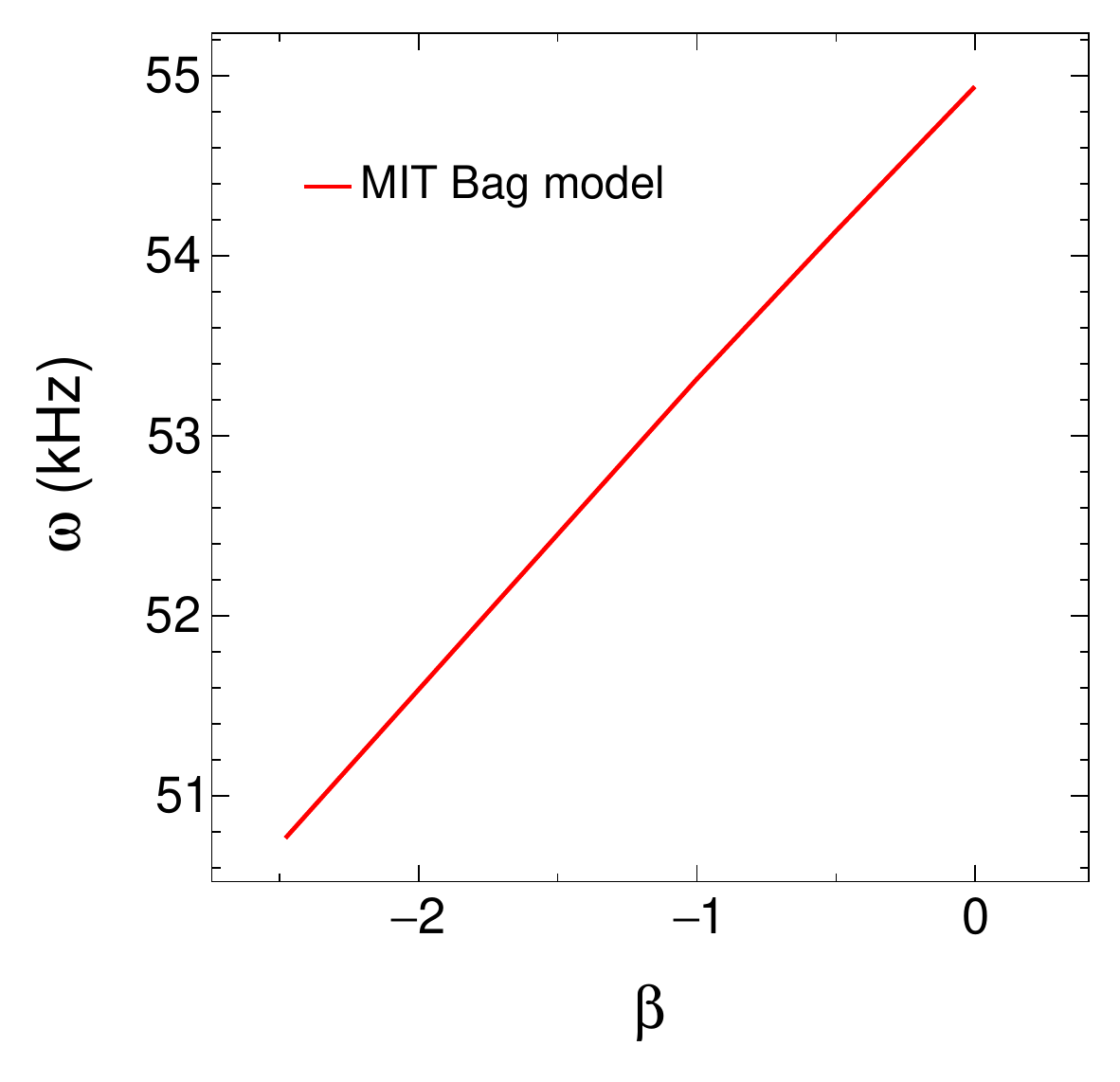}}
        \vspace{-0.3cm}
        \caption{First panel: Variation of mass with radius of strange stars 
for the MIT Bag model EoS. Second panel: Variation of GWE frequencies within 
the allowed range of $\beta$ for the MIT Bag model EoS. Here (and also in 
rest of the cases) the Bag constant $B =$ ($168$ MeV)$^4$ is used.}
        \label{fig1}
        \end{figure*}

\begin{figure*}[!h]
        \centerline{
        \includegraphics[scale = 0.32]{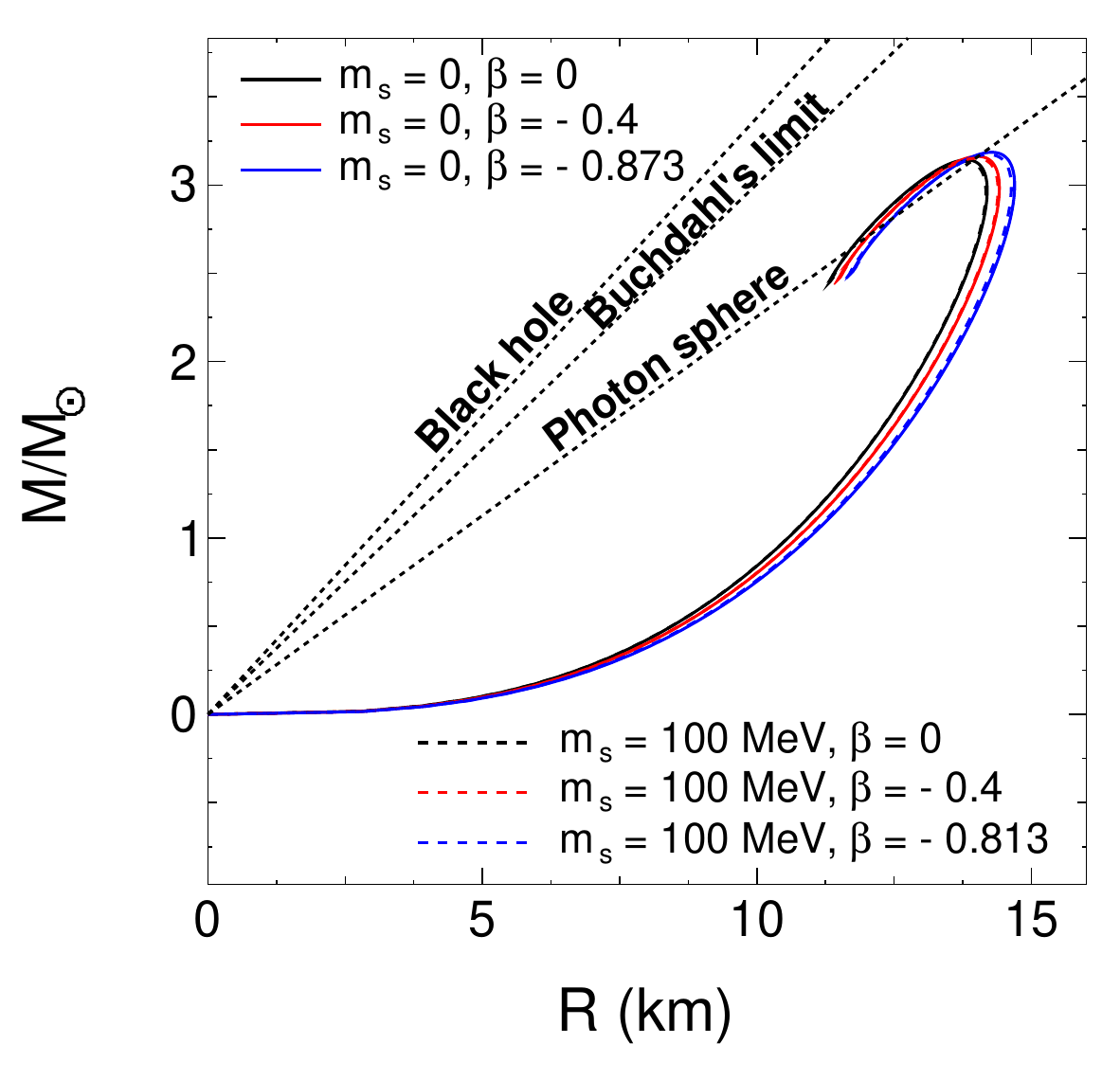}\hspace{1cm}
        \includegraphics[scale = 0.32]{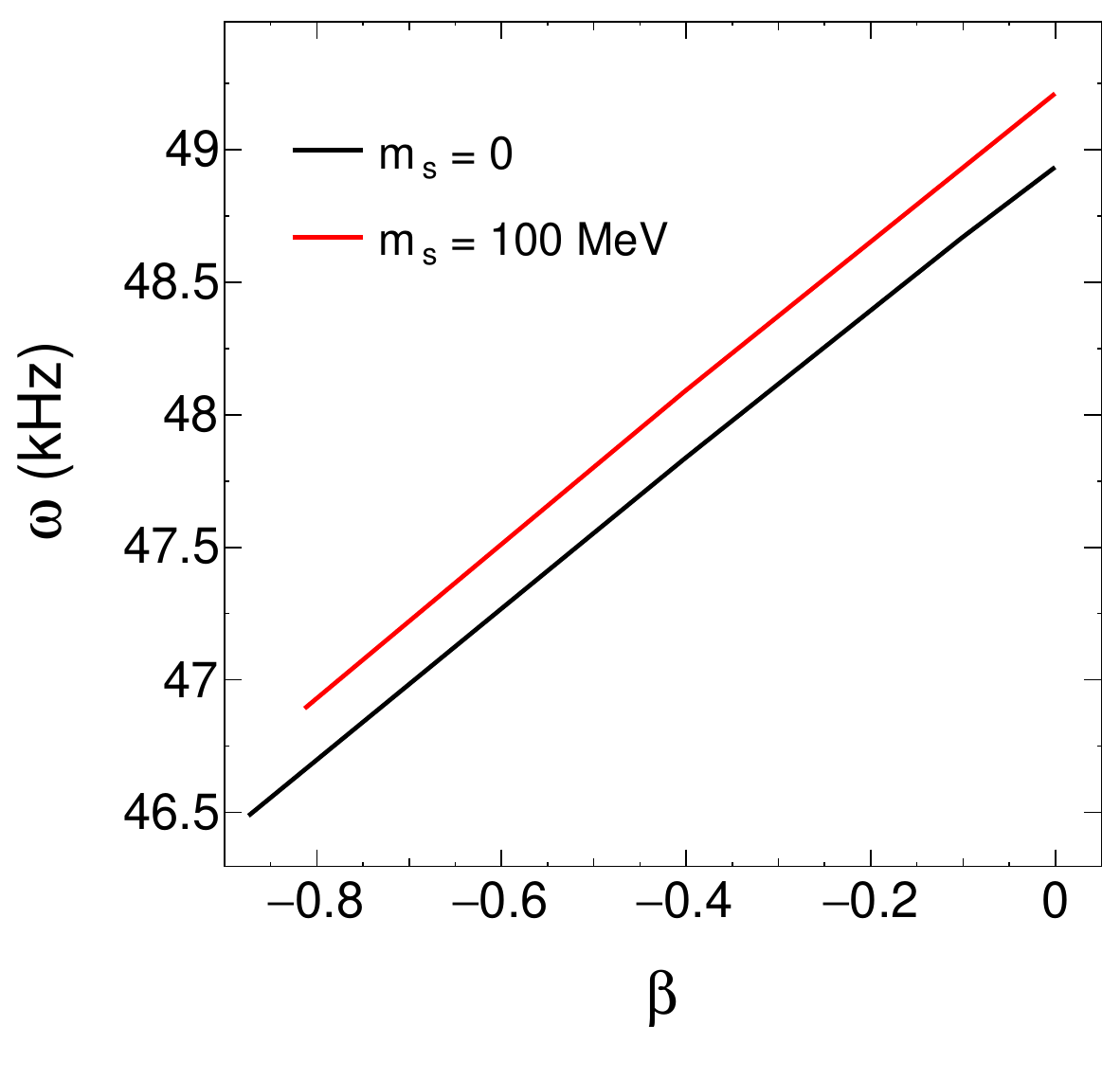}}
        \vspace{-0.3cm}
        \caption{First panel: Variation of mass with radius of strange stars 
for the CFL phase state with $m_s=0$ and $m_s=100$ MeV, and with different 
$\beta$ values. Second panel: Variation of GWE frequencies within allowed 
range of $\beta$ for the CFL state. Here $\Delta=350$ MeV is used.}
        \label{fig2}
        \end{figure*}

\begin{figure*}[!h]
        \centerline{
        \includegraphics[scale = 0.32]{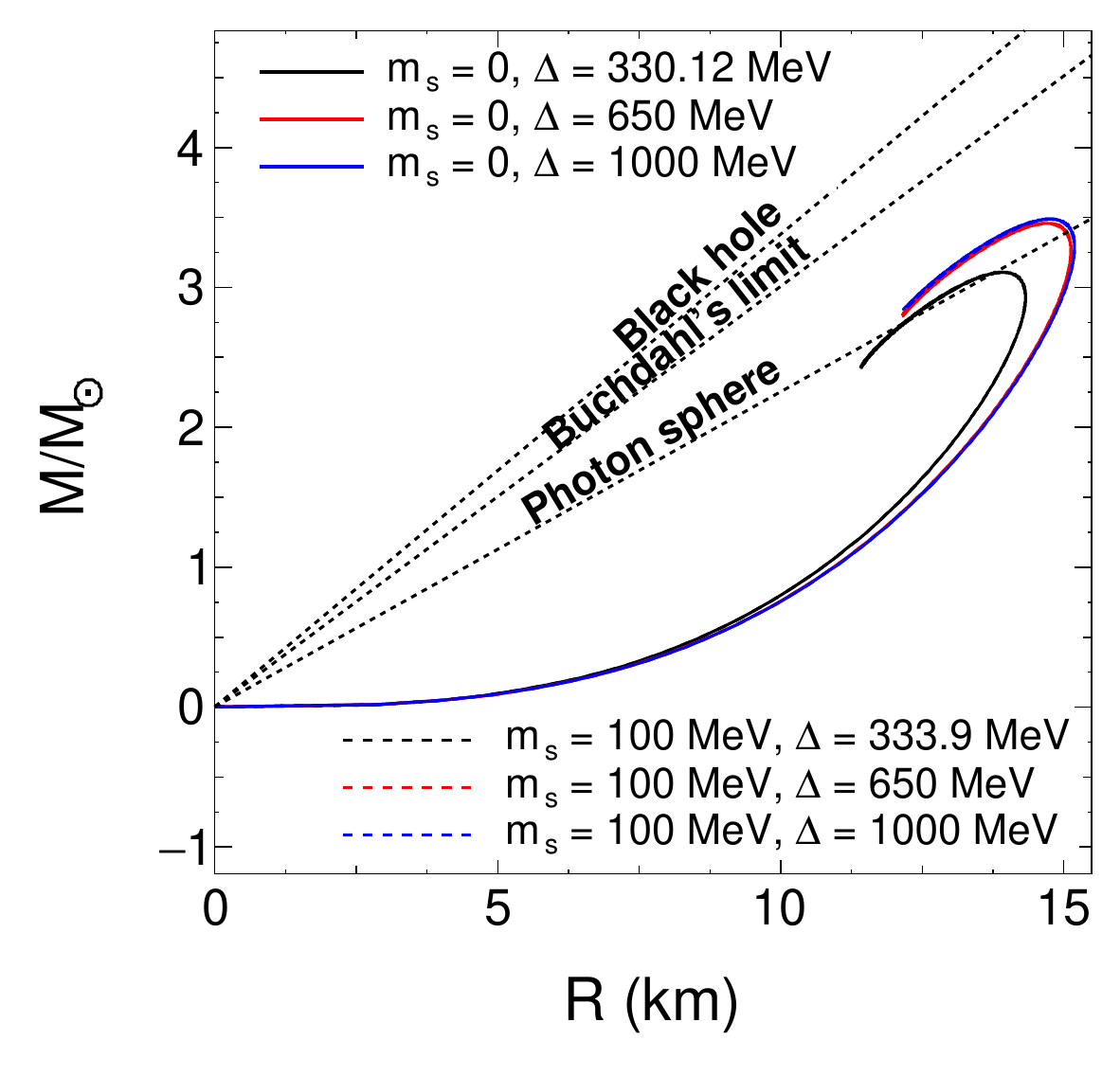}\hspace{1cm}
        \includegraphics[scale = 0.32]{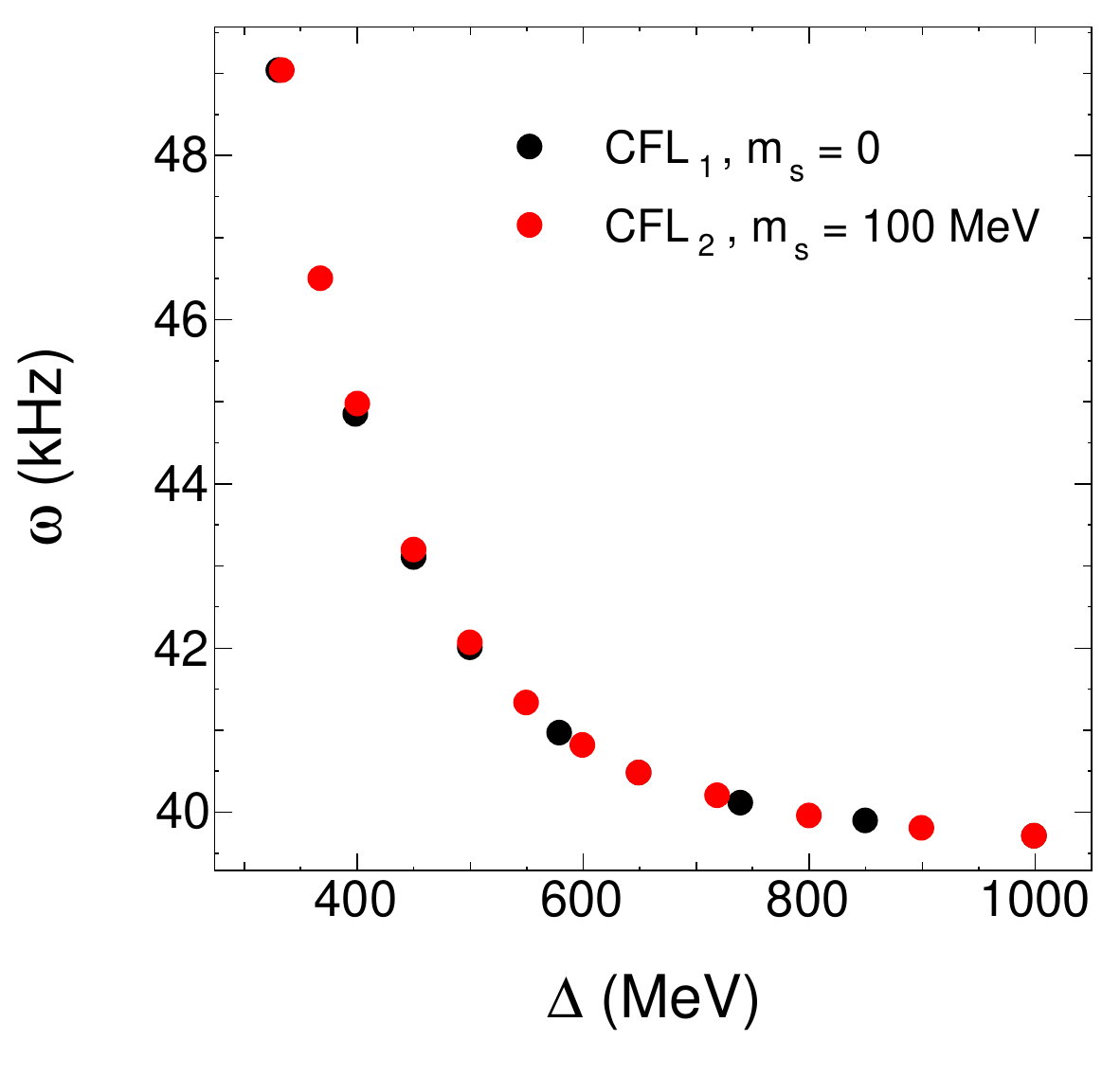}}
        \vspace{-0.3cm}
        \caption{First panel: Variation of mass with radius of strange stars 
with the MIT Bag model EoS. Second panel: Variation of GWE frequencies with
respect to the energy gap $\Delta$. Here the value of $\beta=-0.5$ is used.}
        \label{fig3}
        \end{figure*}

The solutions of TOV equations for mass and radius of compact stars in MIT Bag 
model and CFL phase state are shown in first panels of Figs.\ \ref{fig1}, 
\ref{fig2} and \ref{fig3} for different values of the parameters of the models.
In these plots together with the M-R curves, the photon sphere limit, 
Buchdahl's limit (considering the GR case) and black hole limit are shown. 
For the case of the MIT Bag model the value of Bag constant $B$ considered here 
is ($168$ MeV)$^4$. This value of the bag constant is considered here because 
it lies well within its accepted range \cite{jb1}. Varying the value of the 
$f(\mathcal{R},T)$ model parameter $\beta$, the mass variation (in units of 
solar mass $M_{\odot}$) with the respective radius (in km) is shown in the 
first panel of Fig.\ \ref{fig1}. The minimum value of $\beta$ for which the 
stellar structure will feature a photon sphere in this case is $\beta=-2.474$. 
Moreover $\beta=0$ will correspond to the GR case. We have found that with 
an increase in $\beta$ value the compactness of the most stable structure in 
the M-R curve increases noticeably. As different $\beta$ values depict stellar 
structures with different masses, radii, compactnesses and hence the echoes of 
GWs emanating from the most stable stars' surfaces will result in different 
frequencies. The variation of the echo frequencies within the constraint 
range of $\beta$ can be visualized from the second panel of Fig.\ \ref{fig1}. A 
linear dependency of GWE frequency with $\beta$ is observed. More detailed 
values of some physical parameters of strange star configurations can be found 
in the Table.\ \ref{tab1}.
\begin{table*}[!h]
\caption{\label{tab1} Parameters of strange stars for the MIT Bag model EoS
with the Bag constant $B =$ ($168$ MeV)$^4$.}\vspace{2mm}
\begin{tabular}{c|c|c|c|c|c}\hline
$\beta$ & Radius (R) & Mass (M) & Compactness  & GWE & Surface \\[-2pt] 
        & (in $\mbox{km}$) & (in $M_{\odot}$) & (M/R) & frequency (kHz)& redshift (Z)\\ \hline 
-2.474  & 11.28  & 2.52  & 0.330 & 50.75 & 0.73   \\
-1      & 10.64  & 2.49  & 0.345 & 53.30 & 0.82 \\
0       & 10.26  & 2.46  & 0.354 & 54.94 & 0.89 \\ \hline
\end{tabular}
\end{table*}       

The M-R curves for the strange stars with the CFL phase EoS are shown in the 
left panels of Figs.\ \ref{fig2} and \ref{fig3}. For the CFL phase state we 
have two free parameters, the quark mass $m_s$ and the pairing energy gap 
$\Delta$. With $\Delta=350$ MeV, the M-R curves are showing for different 
$\beta$ values and for two considered mass cases as $m_s= 0$ and $m_s=100$ MeV 
in the first panel of Fig.\ \ref{fig2}. For the first case i.e., $m_s=0$ case 
the $\beta$ value corresponding to the lower limit on compactness is found to
be $-0.873$. Thus stellar structures corresponding to $\beta\geq -0.873$ are 
eligible candidates to have a photon sphere. As in the case of the MIT Bag 
model, stars corresponding to GR cases are more compact in nature as compared 
to that in $f(\mathcal{R},T)$ gravity cases. Again, the variation of GWE 
frequencies with the allowed range of $\beta$ are shown in the second panel of 
Fig.\ \ref{fig2}. Monotonically increasing frequencies with increase in 
$\beta$ values are obtained. On the other hand for $m_s=100$ MeV the lower 
limit of compactness is found to be $-0.813$. As compactness is slightly 
larger for the massless quark case than the massive quark case, the echo 
frequencies are also slightly higher for the massless quark case $m_s=0$ than 
that for the massive quark case $m_s=100$ MeV. Table.\ \ref{tab2} and 
\ref{tab3} show distinct values of the different physical parameters of the 
compact stars under study in these cases.
\begin{table*}[!h]
\caption{\label{tab2} Parameters of strange stars for the CFL phase with 
$m_s=0$ and $\Delta=350$ MeV.}\vspace{2mm}
\begin{tabular}{c|c|c|c|c}\hline
$\beta$ & Radius (R) & Mass (M) & Compactness  & GWE\\[-2pt] 
        & (in $\mbox{km}$) & (in $M_{\odot}$) & (M/R) & frequency (kHz)\\ \hline 
-0.873    & 14.28  & 3.19  & 0.330 & 46.48    \\
-0.40      & 14.02  & 3.16  & 0.333 & 47.33  \\
0         & 13.82  & 3.14  & 0.336 & 48.93 \\ \hline
\end{tabular}
\end{table*}
\begin{table*}[!h]
\caption{\label{tab3} Parameters of strange stars for the CFL phase with 
$m_s=100$ and $\Delta=350$ MeV.}\vspace{2mm}
\begin{tabular}{c|c|c|c|c|c}\hline
$\beta$ & Radius (R) & Mass (M) & Compactness  & GWE & Surface\\[-2pt] 
        & (in $\mbox{km}$) & (in $M_{\odot}$) & (M/R) & frequency (kHz)& redshift (Z) \\ \hline 
-0.813    & 14.22  & 3.17  & 0.330 & 46.89 & 0.74  \\
-0.40      & 14.00  & 3.15  & 0.333 & 48.08 & 0.75  \\
0         & 13.80  & 3.13  & 0.336 & 49.21 & 0.78 \\ \hline
\end{tabular}
\end{table*}

Now, choosing a value of $\beta$ in its allowed range as $\beta=-0.5$, the 
significance of pairing energy gap $\Delta$ on mass, radius and hence 
compactness can be visualized from Fig.\ \ref{fig3} and Tables \ref{tab4}, 
\ref{tab5}. In the first panel of Fig.\ \ref{fig3} the M-R curves are shown 
while varying $\Delta$ for $m_s=0$ and $m_s=100$ MeV. Now for the considered 
$\beta$ value the minimum value of $\Delta$ needed to support photon sphere 
is $330.12$ MeV for the case of $m_s=0$ and $333.9$ MeV for the case of 
$m_s=100$ MeV. While approaching to a larger value of $\Delta$ the compactness 
increases vary slightly. Again echo frequencies decrease exponentially with 
increase in $\Delta$ values as shown in the right panel of Fig.\ \ref{fig3}. 
For the two masses, the echo frequencies are found to be nearly equal for all 
the $\Delta$ values.
\begin{table*}[!h]
\caption{\label{tab4} Parameters of strange stars for the CFL phase with 
$m_s=0$ and $\beta=-0.50$.}\vspace{2mm}
\begin{tabular}{c|c|c|c|c}\hline
$\Delta$ & Radius (R) & Mass (M) & Compactness  & GWE\\[-2pt] 
(in MeV) & (in km) & (in $M_{\odot}$) & (M/R) & frequency (kHz)\\ \hline 
330.12   & 13.93 & 3.11  & 0.330 & 49.03    \\
650      & 14.70 & 3.46  & 0.348 & 40.47 \\
1000     & 14.76 & 3.49  & 0.349 & 39.71 \\ \hline
\end{tabular}
\end{table*}
\begin{table*}[!h]
\caption{\label{tab5} Parameters of strange stars for the CFL phase with 
$m_s=100$ and $\beta=-0.50$.}\vspace{2mm}
\begin{tabular}{c|c|c|c|c}\hline
$\Delta$ & Radius (R) & Mass (M) & Compactness  & GWE\\[-2pt] 
(in MeV) & (in km)  & (in $M_{\odot}$) & (M/R) & frequency (kHz)\\ \hline 
333.9    & 13.93  & 3.11  & 0.330 & 49.03   \\
650      & 14.70  & 3.46  & 0.348 & 40.47 \\
1000     & 14.76  & 3.49  & 0.349 & 39.71 \\ \hline
\end{tabular}
\end{table*}


\section{Physical properties of stellar structures}\label{numerical}
Besides the study of M-R curves, which indeed is an important point to study 
the stellar structures, there are some other physical parameters that need to 
be addressed. So in this section some physical parameters of strange star 
configurations viz., metric potential, surface redshift, adiabatic index are 
discussed briefly. 

\begin{figure*}[!h]
        \centerline{
        \includegraphics[scale = 0.32]{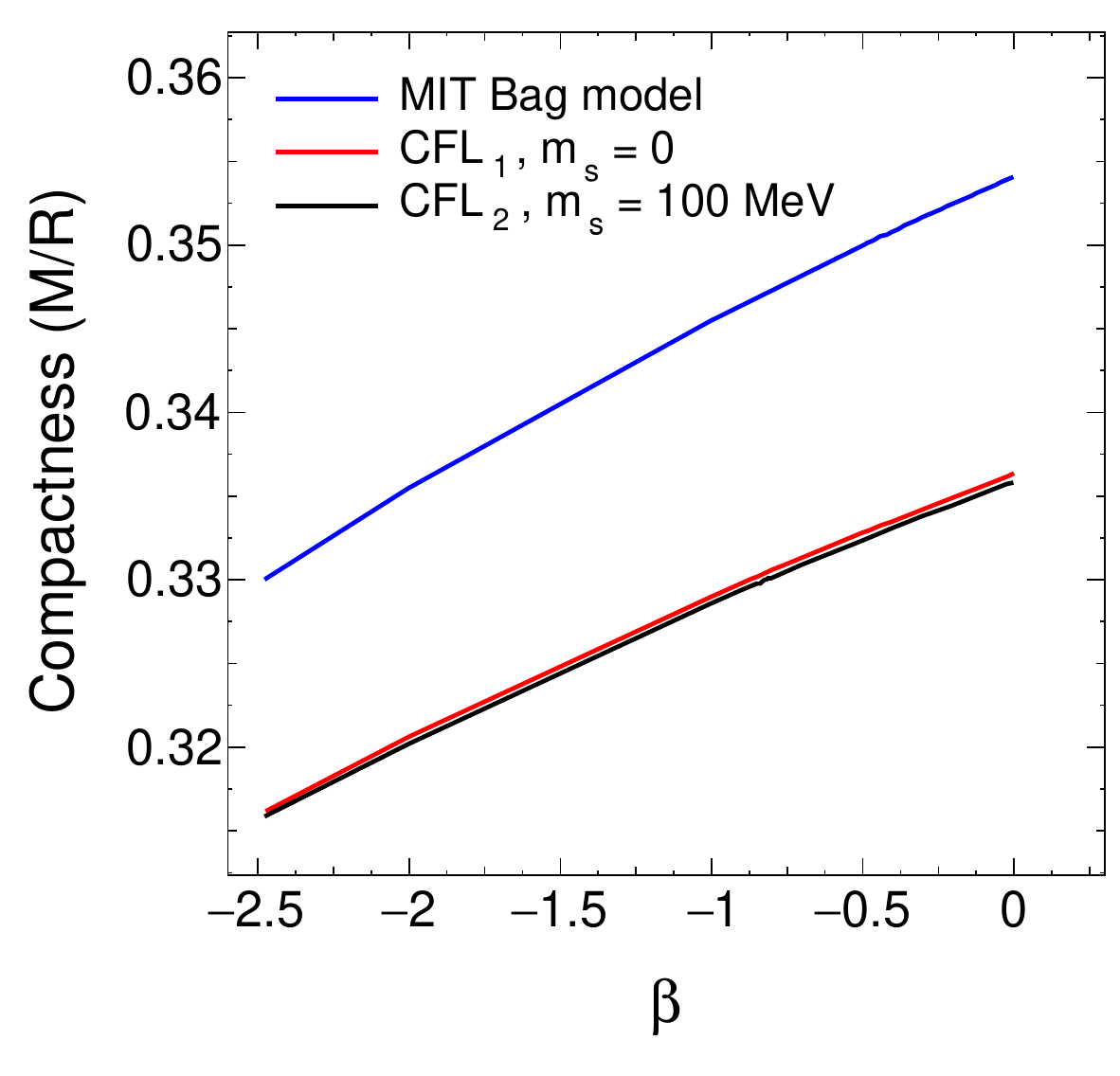}}
        \vspace{-0.3cm}
        \caption{Variation of compactness of stellar structure with the 
        $F(\mathcal{R},T)$ gravity model parameter $\beta$ for the MIT Bag 
model and the CFL phase of quark states. The Bag  constant 
$B = (168\mbox{MeV})^4$ and the energy gap $\Delta = 350$ MeV are used in the
respective plots.}
        \label{fig4}
        \end{figure*}

As mentioned in earlier sections, the present study focuses on two EoSs to 
describe strange quark matter. The first one is the MIT Bag model EoS and the 
other is the interacting quark matter EoS. For these two EoSs with the Bag 
constant $B = (168\mbox{MeV})^4$ the variations of compactness of strange 
stars within a range of $\beta$ values, i.e.~ within a range of values of the 
model parameter of the $f(\mathcal{R},T)$ gravity are shown in 
Fig.\ \ref{fig4}. For all the considered cases of EoSs almost linear 
variations are observed. With a large $\beta$ value the stellar structure with 
a large compactness is obtained. Also the compactness of stars with the MIT 
Bag model EoS along the entire range of $\beta$ values is larger as compared 
to that of CFL phases. The CFL phase with $m_s=0$ is more compact than that 
of the massive case i.e., $m_s=100$ MeV. For these CFL phases the pairing 
energy gap is considered to be $\Delta = 350$ MeV.
        
\begin{figure*}[!h]
        \centerline{
        \includegraphics[scale = 0.32]{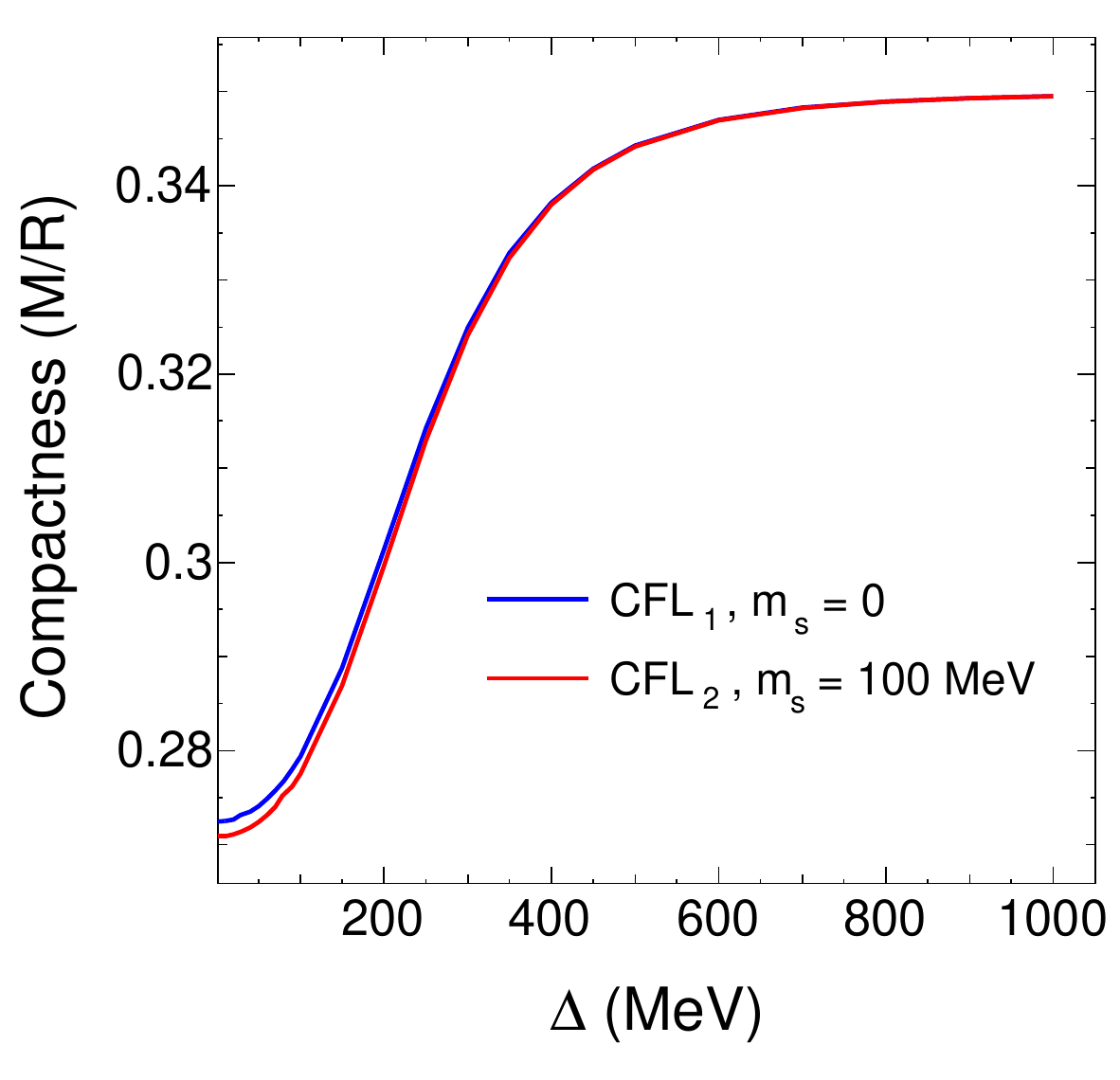}}
        \vspace{-0.3cm}
        \caption{Variation of compactness of stellar structure with pairing 
gap parameter $\Delta$ for the CFL phase of quark states. Here the value of
$\beta$ is taken as $-0.50$.}
        \label{fig5}
        \end{figure*}

The pairing energy plays an important role in the compactness of stellar 
structures. So, the variation of compactness with $\Delta$ is shown in 
Fig.\ \ref{fig5}. We have noticed that for small pairing gaps $\Delta$ 
(lying between 0 to $\sim 600$ MeV) the variation of compactness is rapid. 
After that range the compactness saturates slowly. Also the stellar 
configurations with $\Delta \geq 330$ MeV are found to be compact enough to 
have a photon sphere around their stellar surface. For the maximum $\Delta$ 
value, $1000$ MeV, compact stars with maximum compactness of about $0.349$. 
In Fig.\ \ref{fig5}, for both the mass values, the $f(\mathcal{R},T)$ gravity 
model parameter $\beta$ is taken as $-0.50$.

\begin{figure*}[!h]
        \centerline{
        \includegraphics[scale = 0.32]{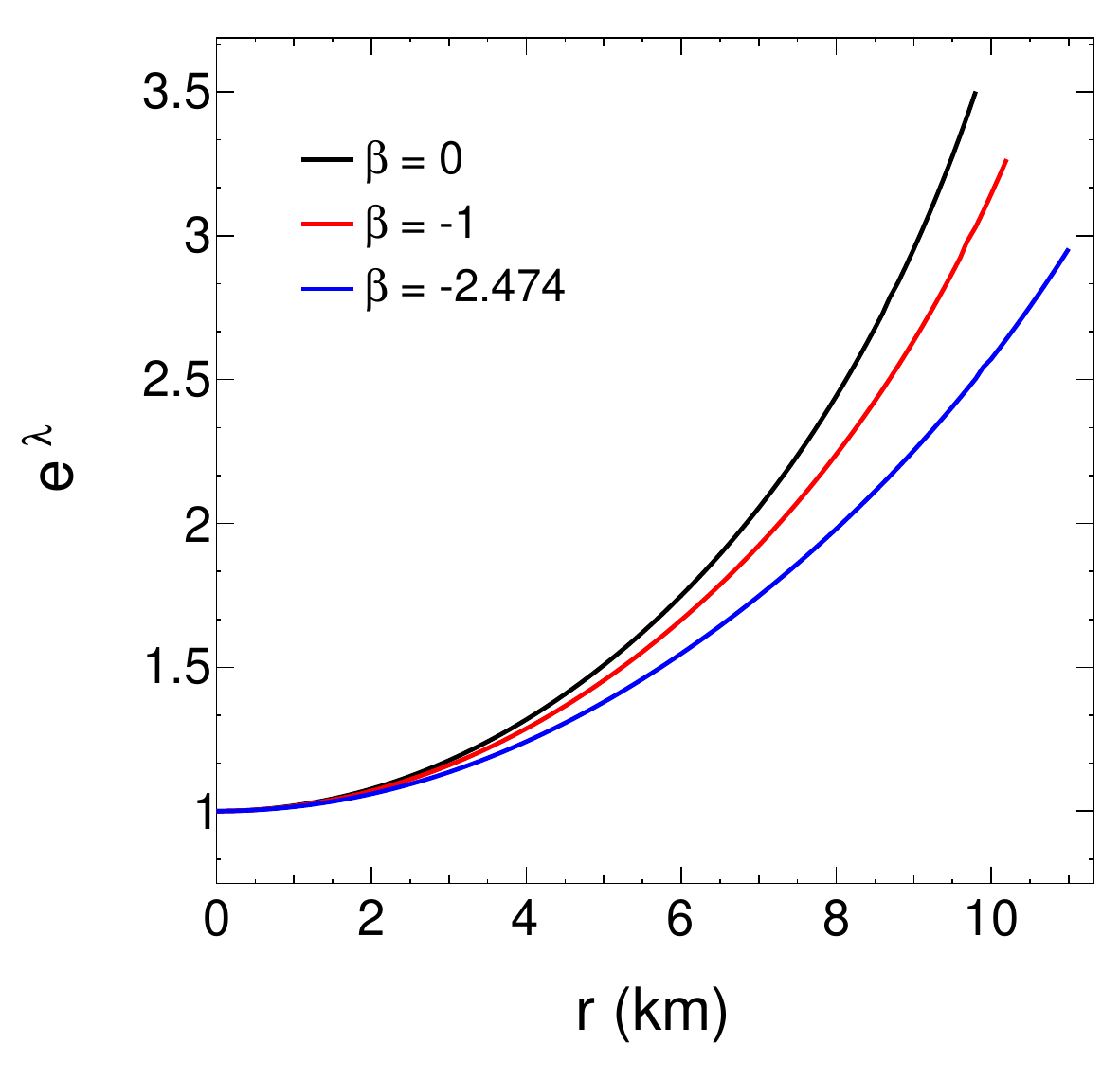}\hspace{1cm}
         \includegraphics[scale = 0.32]{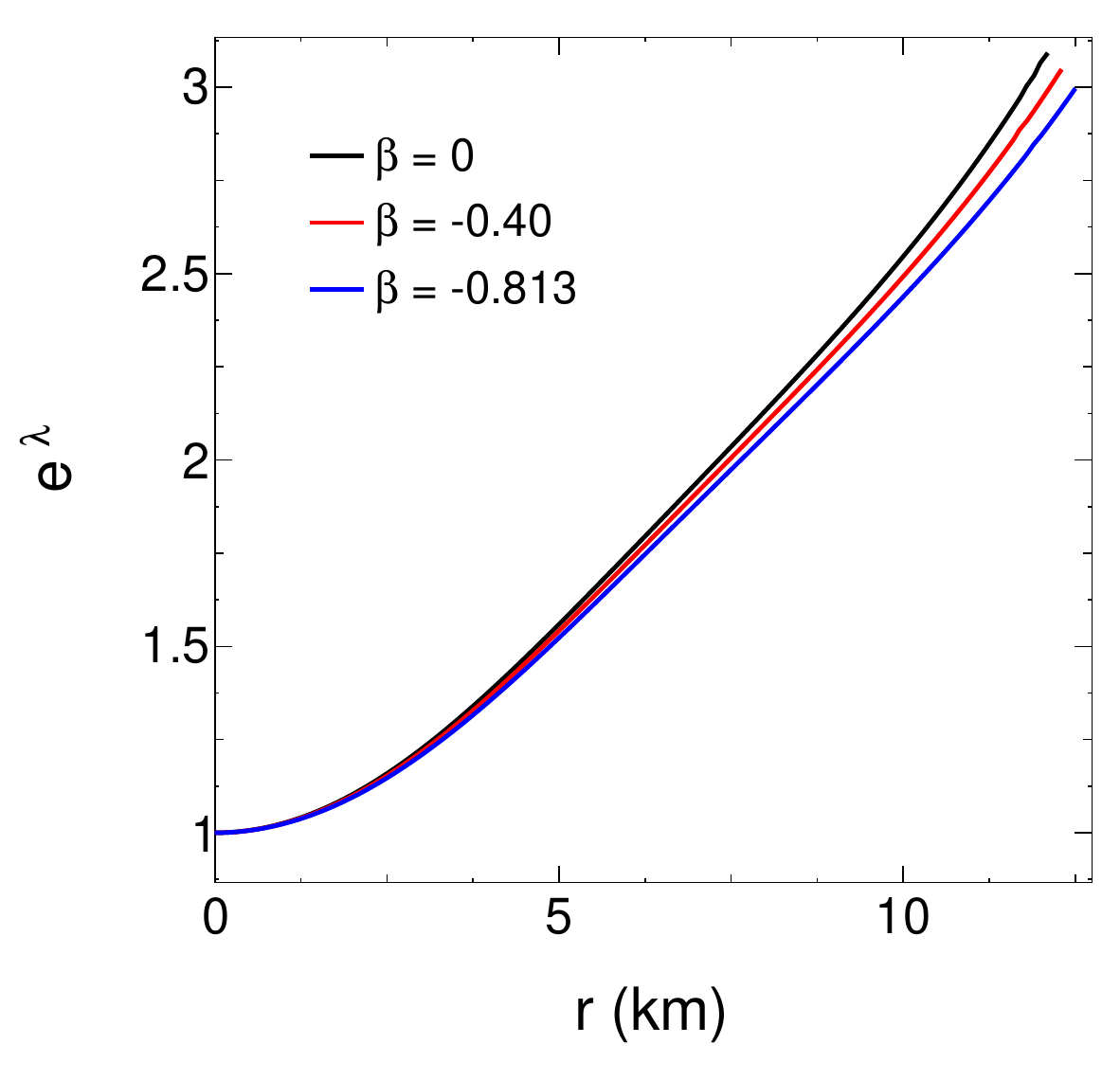}}
        \vspace{-0.3cm}
        \caption{Evolution of metric functions with radial distance $r$ 
for a) MIT Bag model and b) CFL phase states.}
        \label{fig8}
        \end{figure*}

The behaviour of the metric function $e^{\lambda}$ in the stellar object is 
shown in Fig.\ \ref{fig8}. From equation \eqref{h} it is seen that 
$e^{\chi}$, $e^{\lambda} \neq 0$, this indicates that the model is acceptable 
and physically realistic \cite{mustafa}. In the left panel of 
Fig.\ \ref{fig8} the variation of the metric function is shown for the case 
of the MIT Bag model EoS and for the CFL phase EoS with $m_s=100$ MeV it is 
shown in the right panel. It is clear from these two plots that the metric 
function increases monotonically inside the stellar object. 
     
\begin{figure*}[!h]
        \centerline{
        \includegraphics[scale = 0.32]{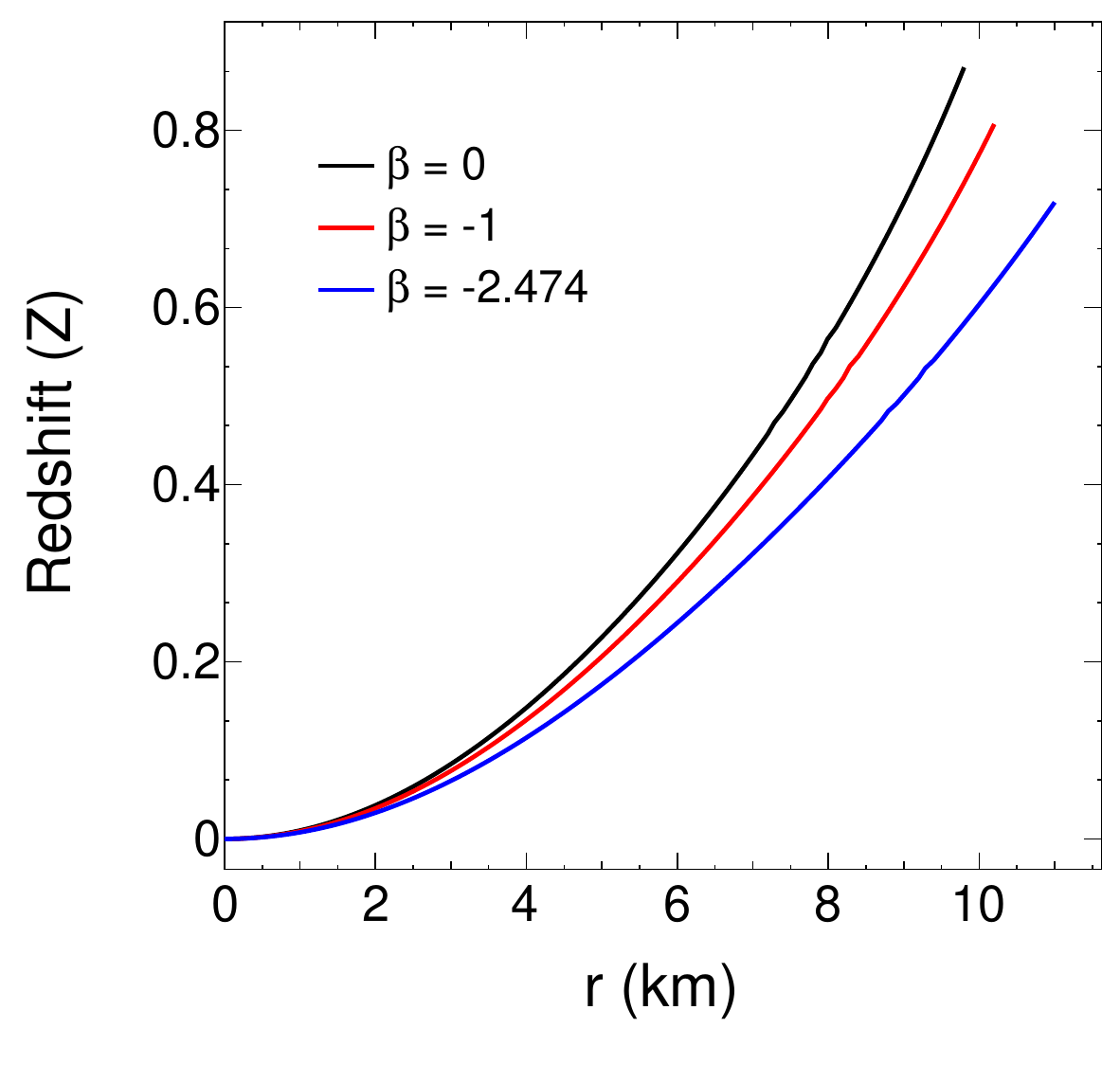}\hspace{1cm}
        \includegraphics[scale = 0.32]{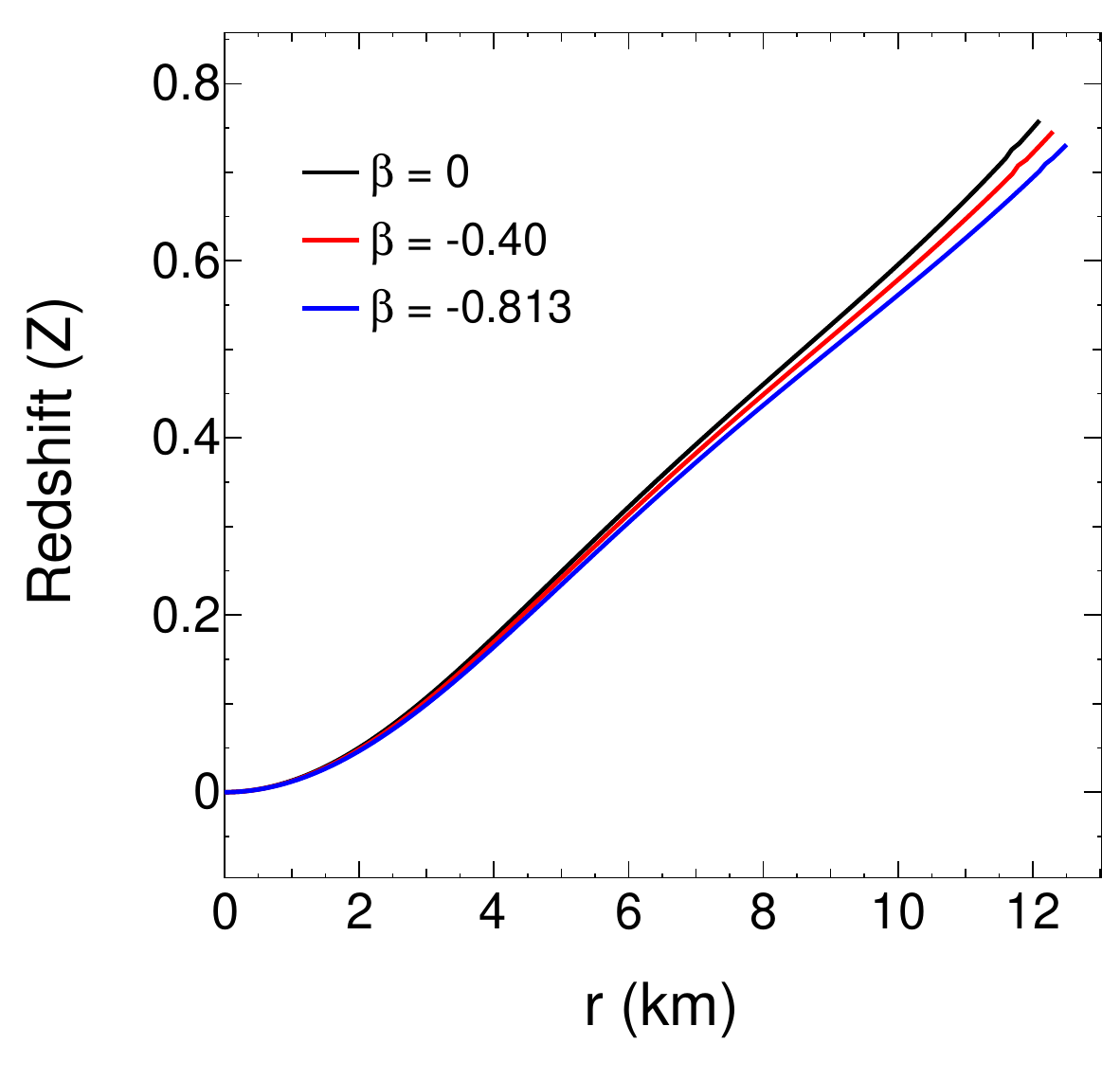}}
        \vspace{-0.3cm}
        \caption{Variation of surface redshift of the stellar structure with 
radial distance $r$ (in km) for a) MIT Bag model and b) CFL phase EoS with 
different $\beta$ values.}
        \label{fig6}
        \end{figure*}

Another important parameter to comment on the stability of relativistic 
objects is the surface redshift. It is related to the compactness of the 
stellar structure and hence it plays an important role in describing 
stability of resulting structures. It describes the relation between the 
interior geometry of the star and its EoS. Here in this study we have considered 
stellar structures which are isotropic, static, spherically symmetric in nature and 
their matter can be described with perfect fluid matter. The compactification 
factor for such a star can be given as
\begin{equation}
\label{z1}
u(r)=\dfrac{m(r)}{r}.
\end{equation}
In terms of this compactification factor the surface redshift of a star can 
be obtained as 
\begin{equation}
\label{a1}
Z=\dfrac{1}{\sqrt{(1-2u)}}-1.
\end{equation}
It is reported that for an isotropic stable stellar configuration the surface 
redshift $Z\leq 2$ \cite{buchdahl, baraco}. For the case of the MIT Bag model 
the variation of redshift with radial distance $r$ is shown in the first 
panel of Fig.\ \ref{fig6}. With the radial distance it increases and attains 
its maximum value at the surface of the star. The values of these surface 
redshifts are listed in Table\ \ref{tab1} for the MIT Bag model EoS. 
It is noted that surface redshift is maximum for $\beta = 0 $ case i.e., the 
for the GR case and minimum for $\beta=-2.474$. Also for all the considered 
$\beta$ maximum redshift values are less than unity and hence eventually 
imply the stability of these structures. 
For the CFL case with $m_s=100$ MeV, the variation of redshift is shown in 
the second panel of Fig.\ \ref{fig6} and the respective values are listed in 
Table \ref{tab3} for the allowed range of $\beta$. Similar to the case of the
MIT Bag model, for this EoS also redshift increases with increase in radial 
distance and redshift is maximum for $\beta=0$. All these redshift values are 
such that they respect the stability criteria of isotropic fluid spheres. 
            
To check a region of stability of relativistic isotropic fluid spheres 
relativistic adiabatic index $\Gamma$ plays an important role. For the 
spherically symmetric spacetime with a perfect fluid the pioneering work on 
instability regimes using this adiabatic index was done by Chandrasekhar 
\cite{Chandrasekhar1964a}. This adiabatic index directly follows from 
the radial perturbations of a spherical star \cite{mtw}. For such stars the 
stability is only ensured when the fundamental mode of radial oscillations is 
a real quantity i.e., $\omega^2>0$ \cite{mtw}. For stability of an isotropic 
sphere $\Gamma > 4/3$ and violation with lower $\Gamma$ values indicates 
instability in the stellar configuration. This adiabatic index can be defined as
\begin{equation}
\label{a2}
\Gamma=\dfrac{p+\rho}{p}\,\dfrac{dp}{d\rho}.
\end{equation}
For a fully Newtonian case, the stability is maintained for 
$\Gamma>4/3$ \cite{mtw, hh}. For the case of relativistic stars it is required 
that $\Gamma>\Gamma_{critical}$ where \cite{mous}, 
\begin{equation}
\Gamma_{critical}=\dfrac{4}{3}+\alpha \dfrac{M}{R}
\end{equation}
with $\alpha$ being a small positive quantity and $M/R$ is the compactness of the 
star.

\begin{figure*}[!h]
        \centerline{
        \includegraphics[scale = 0.32]{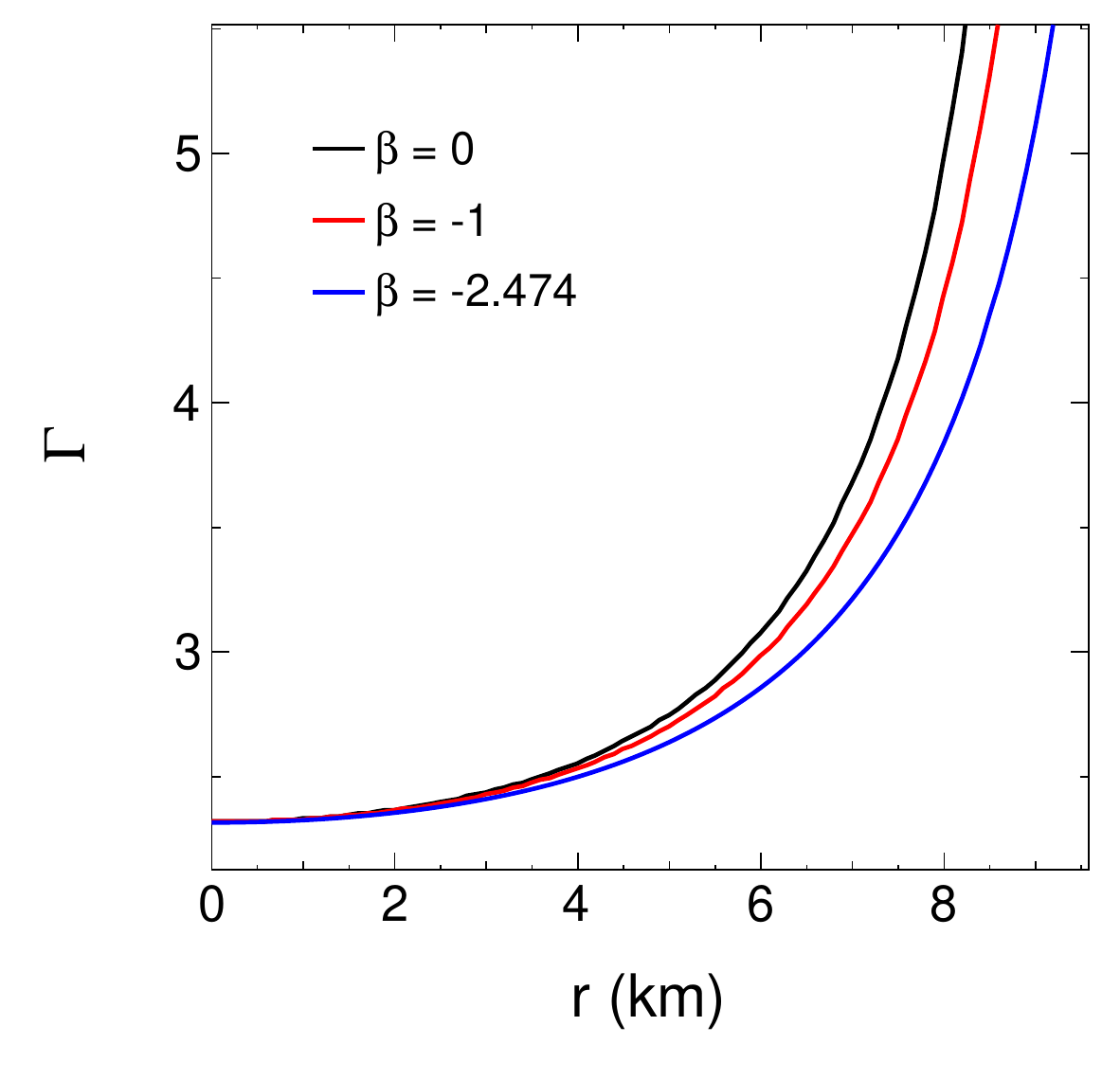}\hspace{1cm}
         \includegraphics[scale = 0.32]{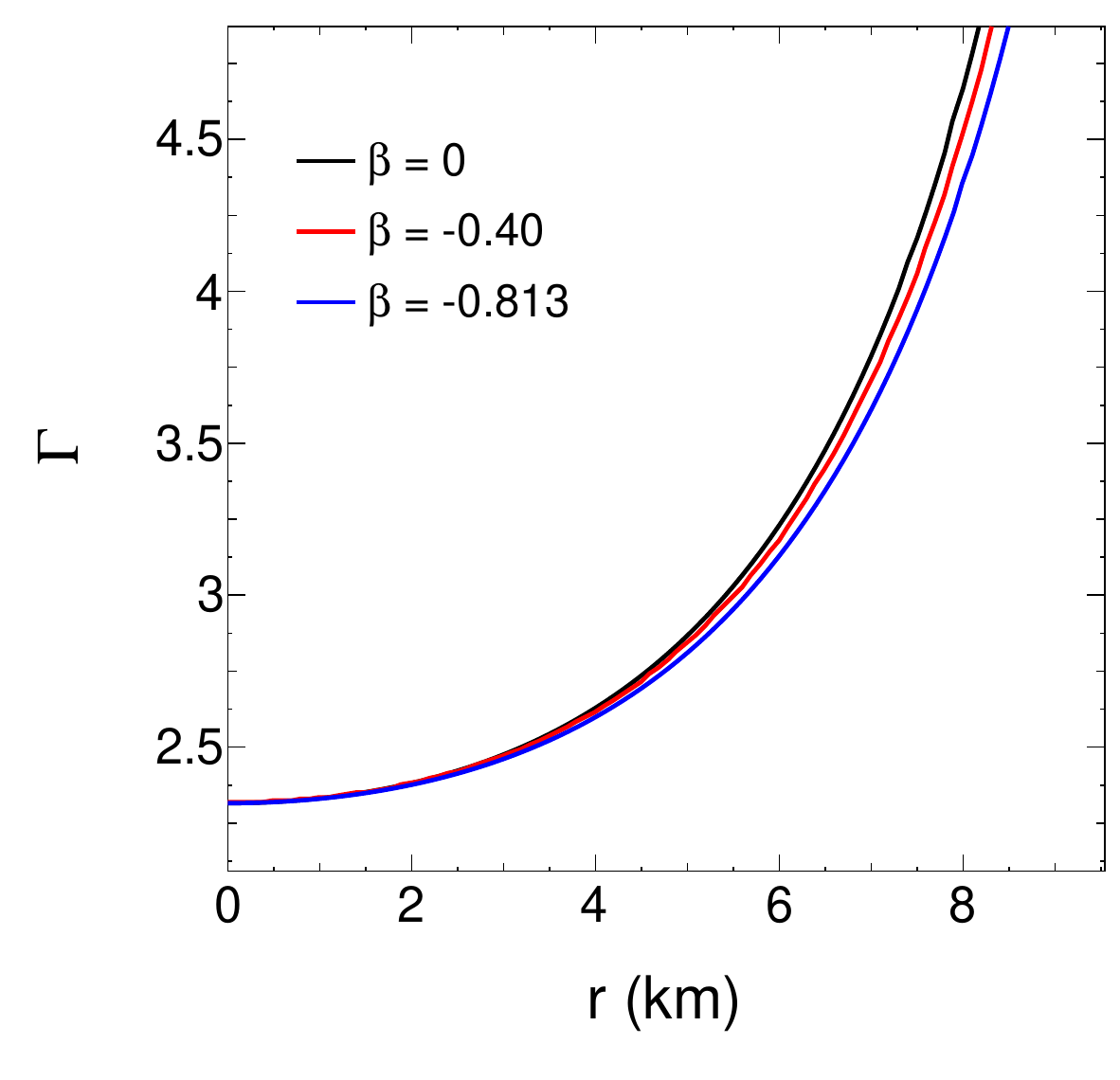}}
        \vspace{-0.3cm}
        \caption{Variation of the adiabatic index with radial distance for 
a) MIT Bag model and b) CFL phase of quark states.}
        \label{fig7}
        \end{figure*}

Now for the considered cases of the present study, the variation of adiabatic 
index $\Gamma$ with the radial distance $r$ is shown in Fig.\ \ref{fig7}. The 
left panel and right panel of this figure corresponds to the MIT Bag model and 
CFL phase of strange quark matter. The adiabatic index is minimum near the 
center of the star. These minimum values for all these cases are greater 
than $4/3$. So it can be inferred that the stellar structures obtained in 
this study are stable in nature. 
       
Furthermore, as discussed in the previous section, for the case of 
ultracompact rotating stars, ergoregion instability and non-linear 
instabilities may occur. In the present study such instabilities can be 
avoided due to the prior consideration of static configuration. However, in a 
realistic scenario, the presence of these compact objects are associated with 
such instabilities.  


\section{Summary and Conclusions}\label{conclusion} 
In this work we have studied the compact stars in the context of 
$f(\mathcal{R},T)$ theory. The $f(\mathcal{R},T)$ gravity model used in this 
study is one of the widely used models viz., 
$f(\mathcal{R},T)=\mathcal{R}+2\beta T$. Using the modified TOV equations 
for this $f(\mathcal{R},T)$ gravity model together with two EoSs, one is the 
MIT Bag model and other is the interacting quark matter EoS, we have obtained 
different stellar solutions. Moreover, for the first time in this work GWE 
frequencies are calculated using realistic CFL induced EoS in the scope of 
$f(\mathcal{R},T)$ gravity. The stability of the obtained solutions are 
also analysed. For the considered cases of EoSs, we have constrained the 
model parameter of the $f(\mathcal{R},T)$ gravity from the perspective of 
compactness. We have found that to feature a photon sphere above the stellar 
surface the minimum value of $\beta$ should be $-2.474$ for the stellar 
structures depicted by the stiffer MIT Bag model. The massless CFL case 
demands a range of $\beta$ as $-0.873<\beta<0$. Whereas for $m_s=100$ MeV, 
the lower limit of $\beta$ changes to $\beta>-0.813$. As discussed earlier 
for the stable equilibrium configurations only negative $\beta$ values are 
allowed. So the constraint $\beta$ values are obtained while respecting 
this stability condition. An important point obtained from this study is 
that it is possible to get compact stars with realistic interacting quark
 matter EoS or the CFL phase state which are able to echo GWs. Moreover, 
beside the stability condition discussed in this work, another important 
condition for stability against radial perturbations of any physical system 
is that the eigenfrequency of the lowest normal mode must be real \cite{hh}. 
For the case of compact star like strange star in 
$f(\mathcal{R},T)=\mathcal{R}+\beta T$ and using MIT Bag model EoS such 
stability was discussed earlier in the Ref.\ \cite{pretel}. 
  
These frequencies of GWEs can play an important role in determining the 
properties of a host of compact stars. The experimental detection of such 
frequency will definitely elucidate the internal composition and physical 
properties of compact stars. The echo frequencies calculated in this study 
are from the theoretical basis. The prediction of echo frequencies will be 
of great use once experimental detection of echo frequencies are possible. At 
present GW detectors like, Advanced LIGO \cite{ligo}, Advanced Virgo 
\cite{virgo} and KAGRA \cite{kagra} are projected at GWs with frequencies of 
$\sim 20$ Hz - $4$ kHz and with amplitudes of $\sim 2\times 10^{-22}$ - $4 \times 10^{-24}$ strain/$\sqrt{\mbox{Hz}}$ \cite{martynov, abbot}. A sensitivity 
of $\ge 2\times 10^{-23}$ strain/$\sqrt{\mbox{Hz}}$ at $3$ kHz are currently 
running at LIGO \cite{currentligo} and Virgo \cite{currentvirgo} 
observatories. Again the proposed third generation detectors like Cosmic 
Explorer (CE) \cite{abbott} and Einstein Telescope (ET) \cite{punturo} with 
optimal arm length of $\approx 20$ km would have the sensitivity to detect 
neutron star oscillations \cite{punturo}. The CE may have sensitivity below 
$10^{-25}$ strain/$\sqrt{\mbox{Hz}}$ at above a few kHz frequencies. Whereas, 
ET will be able to reach the sensitivity of $> 3\times10^{-25}$ 
strain/$\sqrt{\mbox{Hz}}$ at $100$ Hz and of $\sim 6\times10^{-24}$ 
strain/$\sqrt{\mbox{Hz}}$ at $\sim 10$ kHz. Using some enhancement techniques 
of the sensitivity of these present and near future detectors it would be 
possible to detect such frequencies of GWs \cite{martynov,danilishin}. So we 
are optimistic for the detection of such frequencies and thereafter to 
resolve all the mysteries of the interior of compact stars. In this regard such 
theoretical prediction will be very handy as it will lead one to know the 
accurate EoS and hence the properties of the star.

Finally, it needs to be mentioned that in this present work, the echoes 
from compact stellar objects are studied through the echo timescale between 
successive excitations of the photon sphere and the stellar surface. It is a 
very initial technique to analyse echoes of GWs. In this regard the time 
evolutions techniques are required in order to evolve perturbations on such 
configurations. However keeping in view of the context of the 
present work, we have kept it as a possible future direction of the study.

\section*{Acknowledgments}
UDG is thankful to the Inter-University Centre for Astronomy and Astrophysics
(IUCAA), Pune, India for the Visiting Associateship of the institute.

\bibliographystyle{apsrev}
\end{document}